\documentclass[12pt,preprint]{aastex}
\usepackage{natbib}
\def\xmm{\emph{XMM-Newton}}
\def\sax{\emph{BeppoSAX}}
\def\NH{\ifmmode{N_{\rm H}}\else{$N_{\rm H}$}\fi}
\newcommand{\fu}{erg~cm$^{-2}$~s$^{-1}$}
\newcommand{\flu}{erg~cm$^{-2}$}

\def\fs{\hbox{$.\!\!^{\rm s}$}}

\def\farcs{\hbox{$.\!\!^{\prime\prime}$}}

\newcommand{\csq}{\chi^2}
\newcommand{\chisq}{\csq}
\newcommand{\csqn}{\chi^2_\nu}
\newcommand{\chisqn}{\csqn}
\newcommand{\ltsim}{\la}

\begin{document}

\title{Probing the environment in Gamma-ray bursts: the case of
an X-ray precursor, afterglow late onset and wind vs constant density
profile in GRB011121 and GRB011211}

\author{L. Piro\altaffilmark{1},
 M. De Pasquale\altaffilmark{1},
 P. Soffitta\altaffilmark{1},
 D. Lazzati\altaffilmark{2},
 L. Amati\altaffilmark{3},
 E. Costa\altaffilmark{1},
 M. Feroci\altaffilmark{1},
 F. Frontera\altaffilmark{3,4},
 C. Guidorzi\altaffilmark{4},
 J. M. J. in 't Zand\altaffilmark{5},
  E. Montanari\altaffilmark{4},
 L. Nicastro\altaffilmark{6}
}

\keywords{gamma-rays: bursts}

\begin{abstract}
In this paper we present \sax\ and \xmm\ observations of two long
Gamma-ray bursts, the X-ray rich event of December 11, 2001
(GRB011211) and the hard and very bright event of November 21,
2001 (GRB011121). In both events we find evidence of a late X-ray
burst, taking place several minutes after the prompt emission.
In the November burst the spectrum of the X-ray burst  is  much
softer than that of the preceding prompt phase, and consistent
with the spectrum of the afterglow at 1 day. In addition, the tail
of the X-ray burst and  the light curve of the afterglow  at 1 day
are connected by a single power law $\approx(t-t_0)^{-\delta_{\rm
X}}$, when $t_0$ corresponds with the onset of the X-ray burst.
These evidences suggest that the late X-ray burst represents the
onset of the afterglow. A similar conclusion is drawn  for the
December burst. The  temporal and spectral behaviour of the X-ray
and optical afterglows indicate that the fireball evolution in the
December burst takes place in a ISM environment. On the contrary
in the November burst the wind case is revealed by an X-ray decay
slower than that observed in the optical ($\delta_{\rm
X}=1.29\pm0.04$ vs $\delta_{\rm O}=1.66\pm0.06$). The wind profile
should  change into a constant density profile at large radii, in
order to reconcile late-time radio data with a jet. Two other
results are obtained for this burst. An X-ray burst is preceding
by about 30 s the much harder GRB. Contrary to the prediction of
simple models of precursor activity for collapsars, its spectrum
is not consistent with a black body.
Finally, a substantial absorption column
($\NH=(7\pm2)\times10^{22}$~cm$^{-2}$) is detected during the
early part of the prompt emission. This is much greater than that
of the wind, and it is thus likely associated with the region
surrounding the burst.
\end{abstract}

\altaffiltext{1}{Istituto Astrofisica Spaziale \& Fisica Cosmica,
C.N.R., Via Fosso del Cavaliere, 00133 Roma, Italy}

\altaffiltext{2}{Institute of Astronomy, Cambridge, UK}

\altaffiltext{3}{Istituto Astrofisica Spaziale \& Fisica Cosmica,
sezione Bologna, C.N.R., Via Gobetti 101, 40129 Bologna, Italy}

\altaffiltext{4}{Universit\`a di Ferrara, Italy}


\altaffiltext{5}{ Space Research Organization in the Netherlands,
Sorbonnelaan 2, 3584 CA Utrecht, The Netherlands}

\altaffiltext{6}{Istituto Astrofisica Spaziale \& Fisica Cosmica,
sezione Palermo, C.N.R., Via U. La Malfa 153, 90146 Palermo, Italy}

Accepted in Astrophysical Journal

\newpage
\section{Introduction}\label{sec:intro}

Gamma-Ray Bursts (GRBs) and their afterglows are  well explained
by the fireball model, in which a highly relativistic outflow from
the central source produces the observed emission \citep[][and
references therein]{m02}. On the other hand, this process brings
little memory of the central source: the emission of GRB and
afterglow photons by shocks takes place over a distance that is
about 10 orders of magnitude greater than the size of the central
source. In addition, the model is almost independent of the
details of the central source, depending primarily on basic
parameters as the total energy, the collimation angle of the
outflow (jet), the fraction of energy in relativistic electrons
and magnetic fields and the density of the external medium.

Two complementary approaches can be followed  to constrain the
origin of the progenitor. The first is to single out spectral and
temporal features that are produced in the vicinity of the central
source and that can bear direct information on it. This is e.g.
the case of line features
\citep[e.g.][]{pgg+00,lcg_line99,rm_line00,r+02} or SN features
\citep[e.g.][]{gvv+98,paa+00,bkd+99,smg+03,hsm+03}. Shock
break-out at the surface of an exploding massive star are likely
to produce a thermal X-ray precursor. The case of a supernova was
originally explored by \citet{kc78}. More recently several authors
have discussed the case of collimated fireballs erupting from the
stellar surface \citep{mw99,rml02,wm03}.

In the second approach, clues on the progenitor are inferred from
the properties of the environment surrounding the GRB. This is the
case e.g. of measurements of X-ray (and optical) absorption in the
prompt and afterglow phases
\citep[e.g.][]{p04,sfa+04,afv+00,lp02,fac+00}, and the location of
optical transients \citep{bkd02}. Density profiles derived from
afterglow modelling are particularly intriguing, in that the
majority of events are consistent with a constant density
environment, and only in very few cases a wind profile is
preferred \citep[e.g.][]{cl00,pk02}. This is at odd with the
simple expectation of massive star progenitors. Recently
\citet{clf04} proposed a solution to solve this discrepancy,
arguing that a region of constant density would be produced at the
boundary of the wind with the molecular cloud surrounding the
progenitor.

In this paper we present \sax\ \citep{psbsax95,bbpsax+97} and
\xmm\ \citep{jla+01} observations of two long Gamma-ray bursts,
the very bright event GRB011121 and the X-ray rich GRB011211,
hereafter called the November and December burst respectively.
Observations, data analysis and results are presented in
Sect.\ref{sec:obs}.

These events show interesting and peculiar features, bearing
implications on the environment and the progenitor of GRB. In the
November burst we find X-ray bursting preceding by 30 s the hard
prompt event. The origin of this feature is puzzling and it is
discussed in Sect.\ref{sec:precursor} in the framework of fireball
precursors and progenitor precursors.

One intriguing property found in both events is a late bursting in
X-rays taking place several minutes after the GRB trigger. We
discuss this feature in Sect.\ref{sec:rebursting}, showing its
connection with the late afterglow, and arguing that it represents
the onset of the external shock producing the afterglow.

The possible presence of a jet break in the X-ray light curves is
analyzed in Sect.\ref{sec:break}. We then model the afterglow
evolution  taking into account the X-ray measurements to derive
information on the density profile
(Sect.\ref{sec:011121_evolution}). This is particularly important
in the case of the November burst, because this is one of the few
events in which a wind profile is strongly preferred to a constant
density medium \citep{pbr+02}. The other cases include GRB970508
\citep{pk02} and GRB040106 \citep{gpp04} \citep[see
also][]{clf04}.

Broad band modelling of the November burst from radio to X-ray
bands is carried out in Sect.\ref{sec:011121_aft}, where the
discrepancy posed by a jet evolution with the late radio data is
outlined and reconciled in the framework of  a wind termination
shock. The implications of the measurement of X-ray absorption in
this burst are presented and discussed Sect.\ref{sec:absorber}. A
summary of the results and conclusions of the paper is finally
given in Sect.\ref{sec:conclusions}. In this paper we adopt a
cosmology with $\Omega_M=0.3$, $\Omega_{\Lambda}=0.7$ and $H_0=65\
$km~s$^{-1}$~Mpc$^{-1}$. Errors on spectral parameters correspond
to 90\% confidence level ($\Delta\chi^2=2.7$)

\section{Observations}\label{sec:obs}

\subsection{The November burst GRB011121}

\subsubsection{The prompt event: BeppoSAX WFC and GRBM}

\begin{figure}
\centering
\includegraphics[width=0.9\textwidth,origin=c,angle=0]{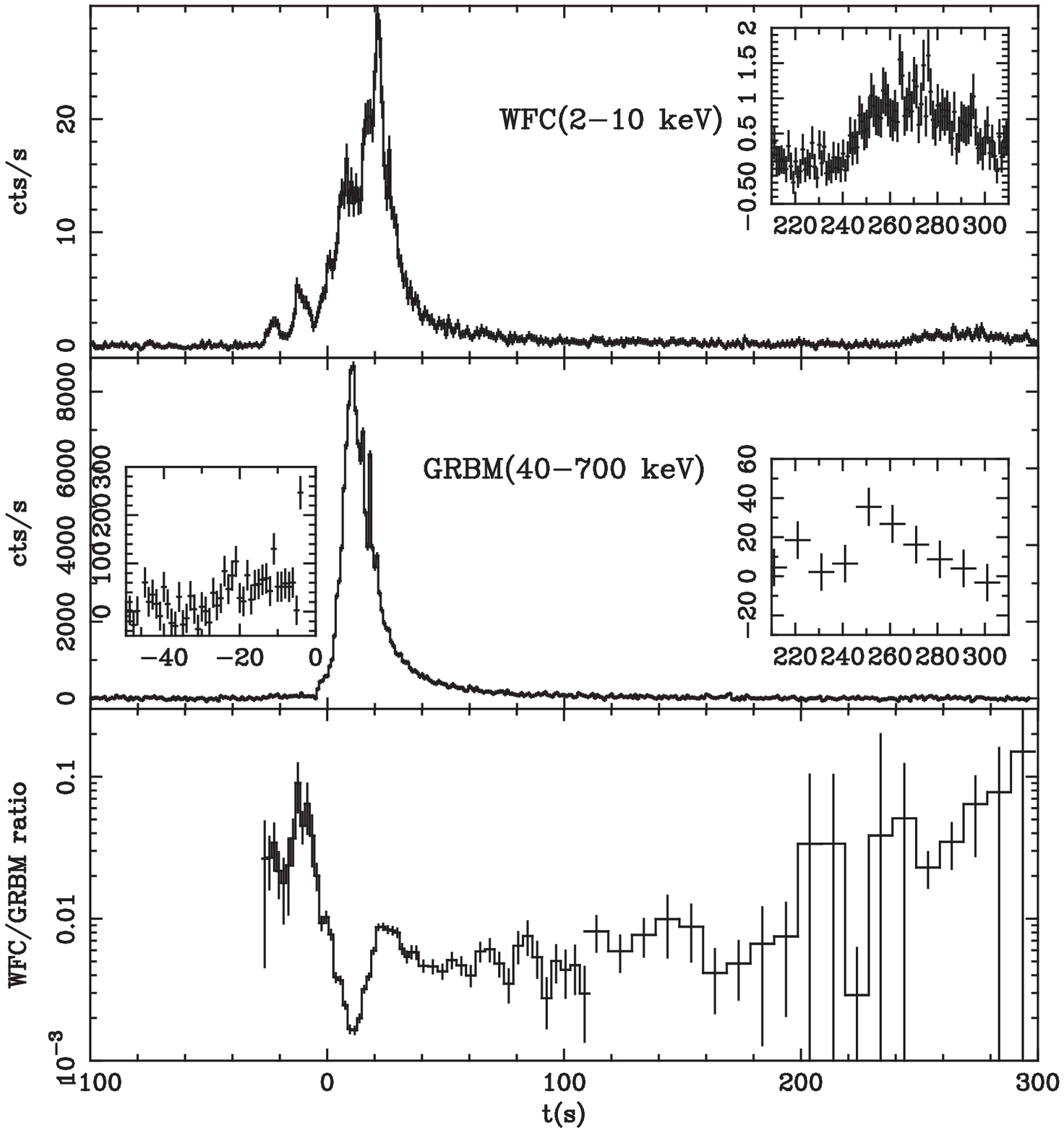}
\caption{The November 21, 2001 burst: \sax\ light curves of the
WFC (upper panel) and GRBM (mid panel). The softness spectral
ratio, given by the WFC/GRBM count rate ratio is shown is the
lower panel (note the log scale). The inset in the mid-left panel
shows  the 40--700 keV corresponding to the X-ray ``precursor".
The right insets in the upper/mid panels  give a blow up of the
light curves of the late X-ray burst.} \label{fig:011121_lc}
\end{figure}

\begin{figure}
\centering
\includegraphics[width=0.8\textwidth,origin=c,angle=270]{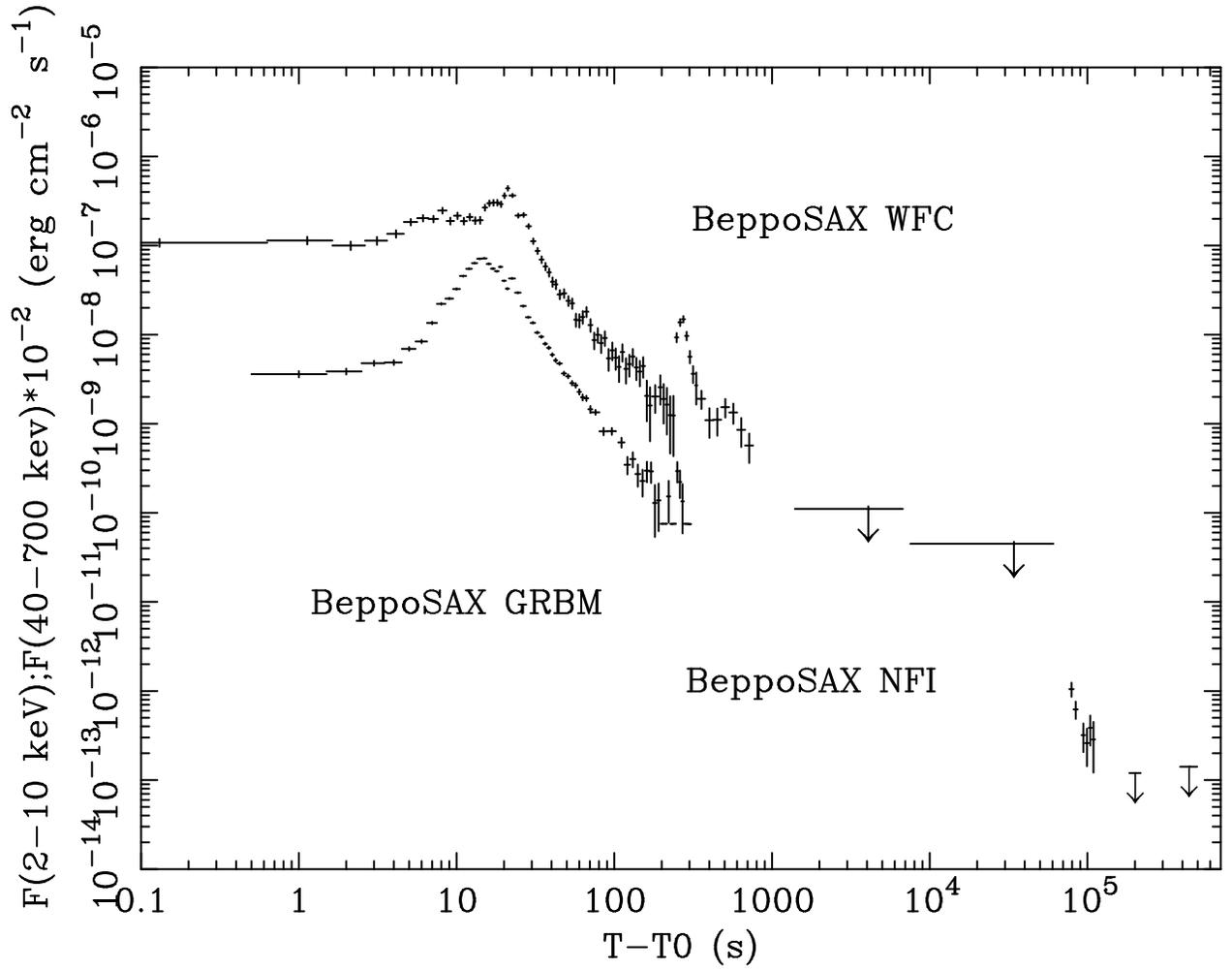}
\caption{The November 21, 2001 burst: \sax\ GRBM (lower data
points, scaled by two orders of magnitude), WFC and NFI (MECS)
light curves.} \label{fig:011121_lc2}
\end{figure}

 GRB011121 was simultaneously detected in
the \sax\ GRBM and WFC  on Nov. 21, 18:47:11.6 UT \citep[][note
the revised trigger time]{piro011121a}. A description of the two
instruments is given in \citet{ffc+97} and \citet{jmb+97}
respectively. In Table~\ref{tab:prompt} we summarize peak fluxes
and fluences in different energy range.
%
The fluence in the 2--700 keV range corresponds to an isotropic
energy of $2.8\times10^{52}$~erg at the redshift of the burst
\citep[$z=0.36$,][see below]{igs+01}. The ratio of X-to-gamma-ray
fluence (peak) are $\frac{S_{2-26}}{S_{40-700}}=0.14$
($\frac{F_{2-26}} {F_{40-700}}=0.08$).  This event was the second
brightest GRB observed by \sax\ in gamma-rays (after GRB990123)
and in X-rays (after GRB010222).  Both the duration of the event
(about 75 s in the GRBM and 120 s in the WFC) and the hardness
ratio are typical of long normal GRB. However, by analyzing the
light curves (Figure \ref{fig:011121_lc} and Figure
\ref{fig:011121_lc2}) we note the following sequence of events:

\begin{itemize}

\item X-ray bursting activity preceding by $\sim30$~s the hard
main pulse of the GRB.  It is composed of two pulses, with a
softness ratio $\approx 20$ times greater than that of the GRB
(lower panel of Figure \ref{fig:011121_lc}). The first pulse is
followed by a second more intense and softer pulse, that decays
just before the onset of the hard GRB. In coincidence with the
X-ray bursting, a faint signal is recorded in the 40--700 keV
range (see left inset of mid panel of Figure \ref{fig:011121_lc});

\item a hard main pulse (-5--30~s) characterized by the typical
behaviour of a GRB, i.e. spectral hardening in the rising part and
hard-to-soft evolution in the decaying part;

\item a hard tail (30--240~s) with a temporal slope of about 1.4;

\item a late X-ray burst (240--310~s) (see upper and mid rightmost
insets of Figure \ref{fig:011121_lc}) with a softness ratio
$\approx 10$ times larger than that of the preceding phase;

\item a slow X-ray tail (310--716 s).

\end{itemize}

\begin{figure}
\centering
\includegraphics[width=0.8\textwidth,origin=c,angle=270]{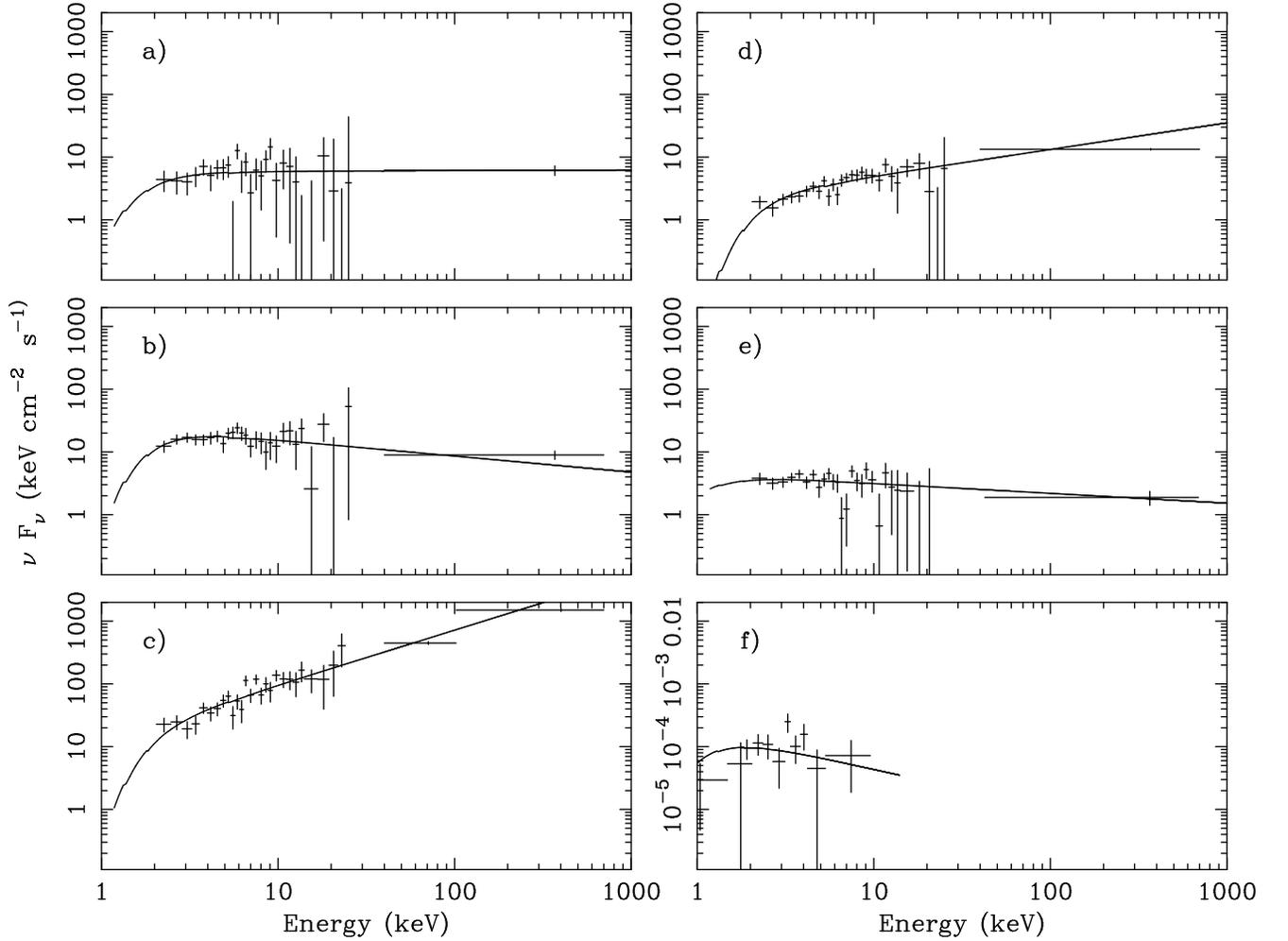}
\caption{The November 21, 2001 burst spectral evolution. We plot a
selection of spectra ($\nu F_{\nu}$) as function of time: Panel
a)``Precursor": Pulse 1. Panel b)``Precursor": Pulse 2. Panel
c):gamma-ray peak. Panel d):slower tail. Panel e):late X-ray
burst. Panel f): afterglow (see Tab\ref{tab:gb011121_spectra}.
Panel a) to e): WFC and GRBM data. Panel f): LECS and MECS. The
continuous line is the best fit absorbed power law model.  The
column density in best fit model of LECS and MECS is equal to the
galactic value. Note, in particular, the sharp transition from a
soft to hard spectrum from the ``precursor" phase to the GRB
phase, an opposite transition to a soft spectrum, when the late
X-ray burst event sets in (panel e) and the similarities of the
spectra of the precursor (panel a,b), the late X-ray burst (panel
e) and the late afterglow (panel f).} \label{fig:011121_spe}
\end{figure}

To study in detail these features, we have  carried out a
time-resolved spectral analysis of the combined WFC and GRBM data.
The results are presented in Table~\ref{tab:gb011121_spectra} and
Figure \ref{fig:011121_spe}.  With regard to the X-ray bursting
preceding the proper GRB and to the late X-ray burst we note the
following.

\begin{itemize}

\item{X-ray bursting preceding the proper GRB}. We have analyzed
separately the spectra of the two pulses. Both are soft, being
fitted by a power law with a energy index $\alpha=1.0\pm0.11$ and
$\alpha=1.25\pm0.10$ for the first and second pulse respectively.
At the onset of the gamma-ray pulse, the spectrum changes abruptly
to $\alpha=0.1$, with the peak of the energy output changing from
the X-ray band to above 1 MeV (Figure \ref{fig:011121_spe}).
Hereafter we will refer to this event as X-ray ``precursor", to
underline its {\it observational} peculiarity with respect to the
hard main pulse of the GRB.  The 2--700 keV fluence is
$2\times10^{-6}$ \flu corresponding to an isotropic energy
$E_{prec}=6\times10^{50}$ erg, i.e. $\approx 2\%$ of the isotropic
energy of the proper GRB.

\item{Late X-ray bursting}. The prompt event is very hard and is
followed by a hard tail with a moderate hard-to-soft evolution
($\alpha\approx0.1-0.6)$ until the late X-ray burst takes place.
In this event, the spectrum  switches to a soft shape
($\alpha=1.15\pm0.15$), keeping this spectral shape in the
subsequent tail and, as shown below, in the follow-up observation
performed around 1 day after the burst. The isotropic energy
contained in this event is $\approx 3\%$ of that of the proper
GRB.

\end{itemize}

Finally we find that the spectra in the early part of the burst,
when fitted with a power law, require  a significant absorption
$\NH=(7\pm2)\times10^{22}$~cm$^{-2}$ at the redshift of the burst.
A fit with the smoothly broken power law proposed by
\citet{bmf+93} gives very low values of $E_0$ (below 10 keV), a
model spectrum rising with energy below $E_0$ and the same
spectral index of the power law fit above $E_0$. Basically, this
model tries to reproduce the curvature due to the low energy
absorption.  In addition, $\chisq$ values are typically worse than
those derived from a simple power law with absorption.

\subsubsection{Follow-up observations: BeppoSAX NFI}

The prompt localization of this burst
\citep{piro011121a,piro011121b} triggered several follow-up
observations, including one by \sax\ itself. This was performed
with  Narrow Field Instruments MECS and LECS starting 21 hours
after the burst. The two instruments are described in
\citet{bccsax+97} and \citet{pmbsax+97} respectively. It was the
first TOO observation performed after a new attitude control mode,
the so-called gyroless mode, was installed in October 2001 and
still being tested. The observation was divided in two parts, the
first covering the period 21--33 hours and the second 52--60 hours
after the burst. The total net exposure time was 33 ks in the MECS
and 9 ks in the LECS. A further observation was carried out 4.5
days after the burst, with a MECS net exposure of 22 ks and 11 ks
in the LECS.

The fading X-ray afterglow of GRB011121 (1SAX J113426$-$7601.4)
was detected in the first part of the first observation with
$F_{\rm 2-10\; keV}=4\times 10^{-13}$ \fu at a position (J2000)
${\rm RA} = 11^{\rm h} 34^{\rm m} 25\fs8$, ${\rm Dec.}=-76\degr
01' 22"$, with an error radius of $50"$ \citep{piro011121c}. This
position is $25"$ away from the optical transient \citep{gsw+03},
well within the error box, verifying the good quality of the
aspect reconstruction in this new pointing mode. In the other
parts of the observation the source was not detected with an upper
limit $\approx 10^{-13}$ \fu. A fit to the light curve with a
power law $t^{-\delta_{\rm X}}$ gives $\delta_{\rm X}=3.8\pm1.9$.
The spectrum is fitted by an absorbed power law with energy
spectral index $\alpha_{\rm X}=1.6\pm0.7$ and column density
consistent with the galactic value
(Table~\ref{tab:gb011121_spectra}).

Observations of the optical counterpart revealed a rather nearby
event \citep[$z=0.36$,][]{igs+01}, and excess emission above the power
law decay, attributed to a SN bump \citep{gsw+03, gks+03}. The
latter authors report a break after 1.3 days, attributed to a
collimated outflows. Radio and optical observations suggested a
fireball in a wind medium \citep{pbr+02}.

\subsection{The December burst GRB011211}

\subsubsection{The prompt event: BeppoSAX GRBM and WFC}
\begin{figure}
\includegraphics[width=0.8\textwidth,origin=c,angle=270]{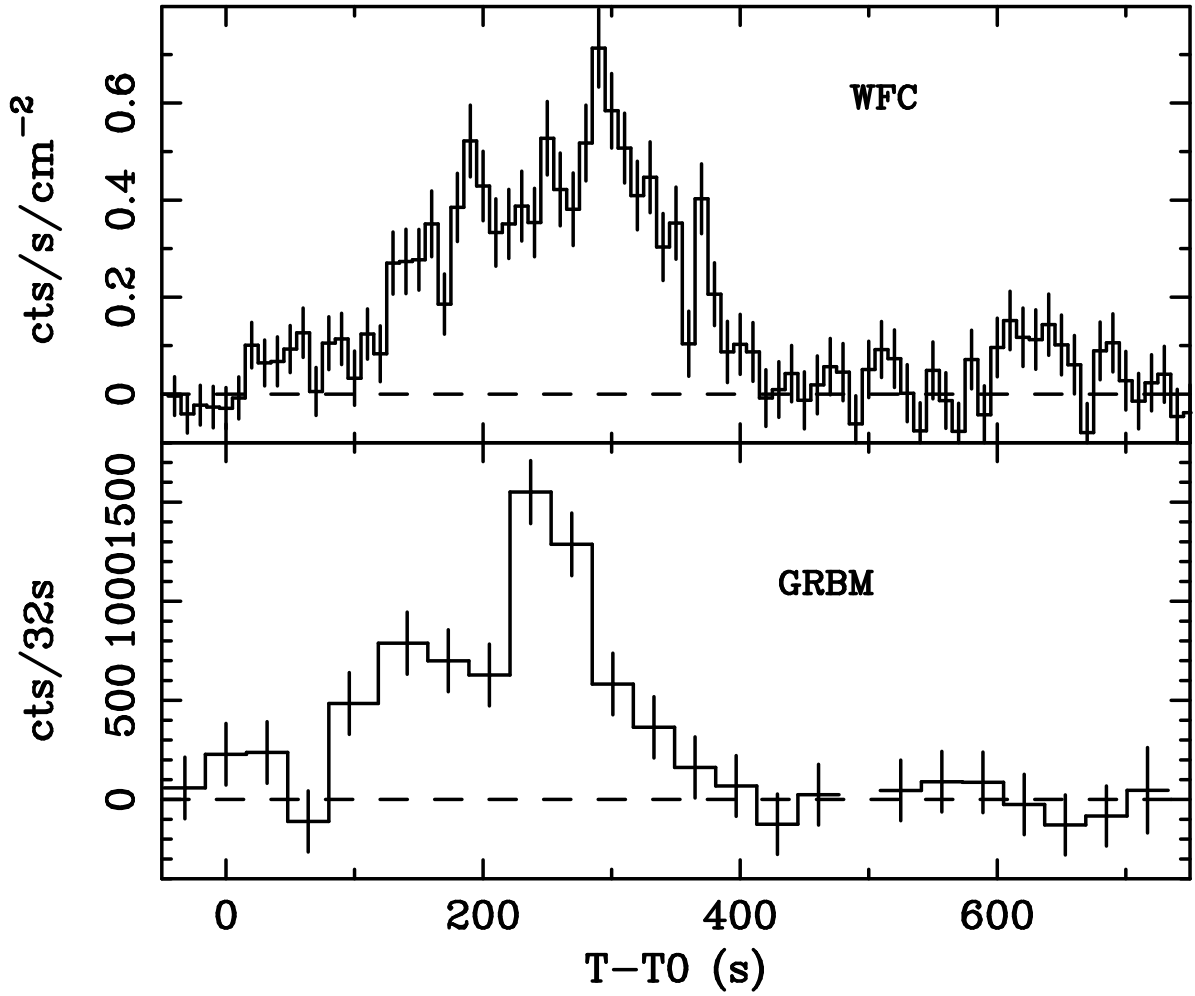}
  \caption{The December 11, 2001 Burst. Light curves  from \sax\
      GRBM (40--700 keV) \& WFC (2--26 keV). Note the late X-ray burst in the WFC data at
      600 s}\label{fig:011211_lc}
\end{figure}

The gamma-ray burst GRB011211 was detected by the \sax\ Wide Field
Camera 1 (WFC: 2--26 keV) on 2001 December 11, 19:09:21 UT
\citep{g+01}. A detailed off-line analysis of the GRBM (40--700
keV) data showed a very shallow and long event \citep{f011211sax},
with a duration of about 270 s, similar to that observed in the
WFC (Figure \ref{fig:011211_lc}).  In Table~\ref{tab:prompt} we
summarize peak fluxes and fluences in different energy ranges.
The fluence in the 2--700 keV range corresponds to a isotropic
energy of $3.6\times10^{52}$~erg at the redshift of the burst
(z=2.14, \citet{fvr+01}, see below). The ratio of X-to-gamma-ray
fluence (peak) are $\frac{S_{2-26}}{S_{40-700}}=0.5$
($\frac{F_{2-26}} {F_{40-700}}=0.3$). These values are higher than
in typical GRBs \citep{fac+00}, classifying the event as an X-ray
rich GRB, i.e. in the region in between ``normal" GRBs and X-ray
flashes. This is also supported by a combined spectral analysis of
the WFC and GRBM spectra of the main prompt pulse (0--400 s) with
the Band model, that gives $E_0=20_{-12}^{+20}$,
$\alpha=0.1_{-0.4}^{+0.2}$, $\beta=1.1\pm0.2$, where $\alpha$ and
$\beta$ are the {\it energy} spectral indices. Though the values
of the spectral indices are consistent with those observed in
normal GRBs \citep[e.g.][]{fac+00} the peak energy
$E_p=E_0(1-\alpha)=18$~keV is substantially lower than the average
value of about 200 keV of the BATSE sample and in the range of
values found in X-ray rich GRBs \citep{kwh+01}.

\begin{figure}
\includegraphics[width=0.8\textwidth,origin=c,angle=270]{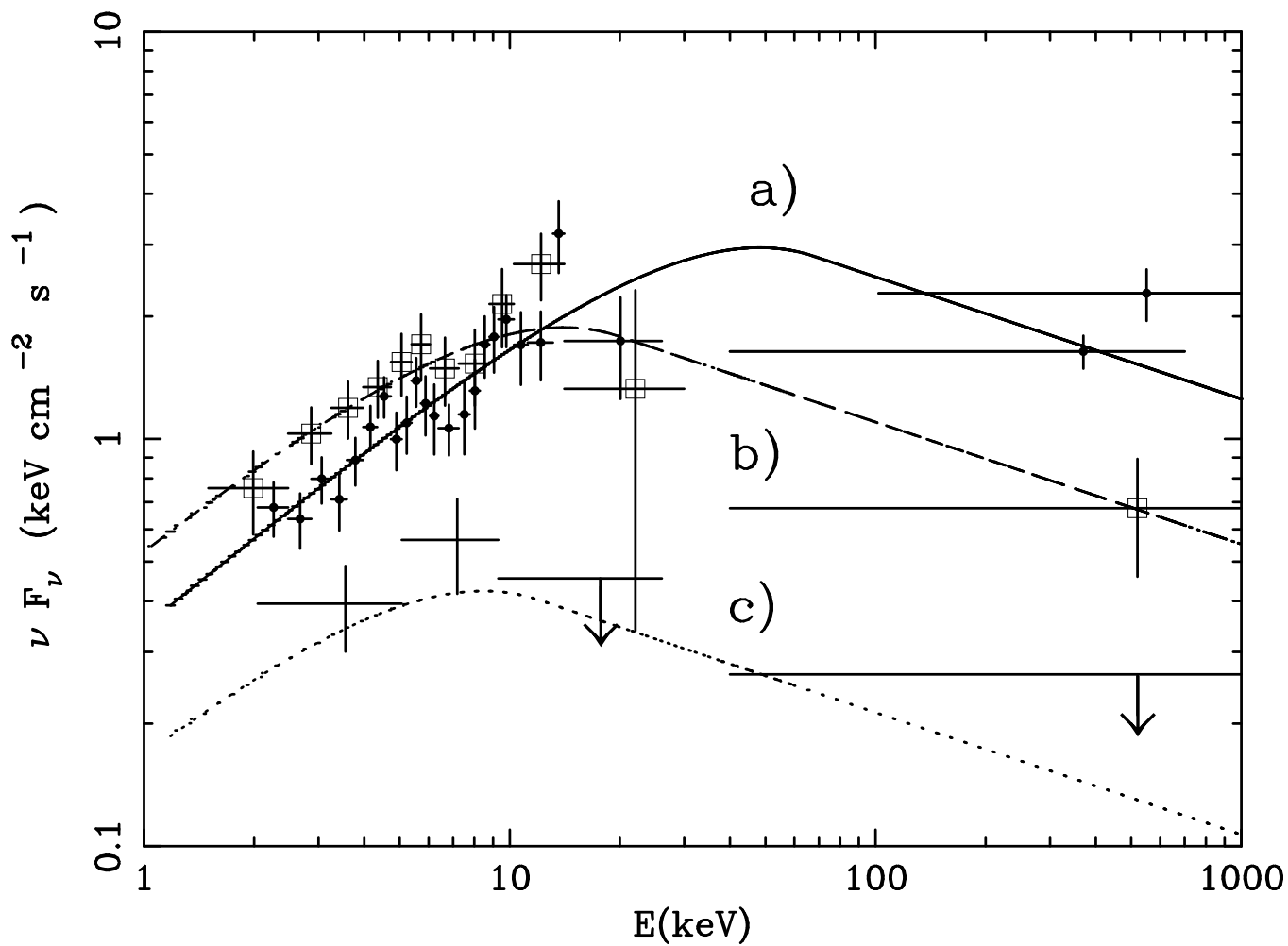}
  \caption{ Spectral  ($\nu F_{\nu}$) evolution of the December 11, 2001 burst
   in the \sax\
      GRBM \& WFC with a Band model in  time slices a) 0--300 s
 (data points: filled circles; model:
      continuous line), b) 300--400 s (data: open squares; model: dashed line)
 and c) 575--675 sec (data: crosses and upper limits; model: dotted line).}
\label{fig:011211_spe}
\end{figure}

\begin{figure}
\includegraphics[width=0.8\textwidth,origin=c,angle=270]{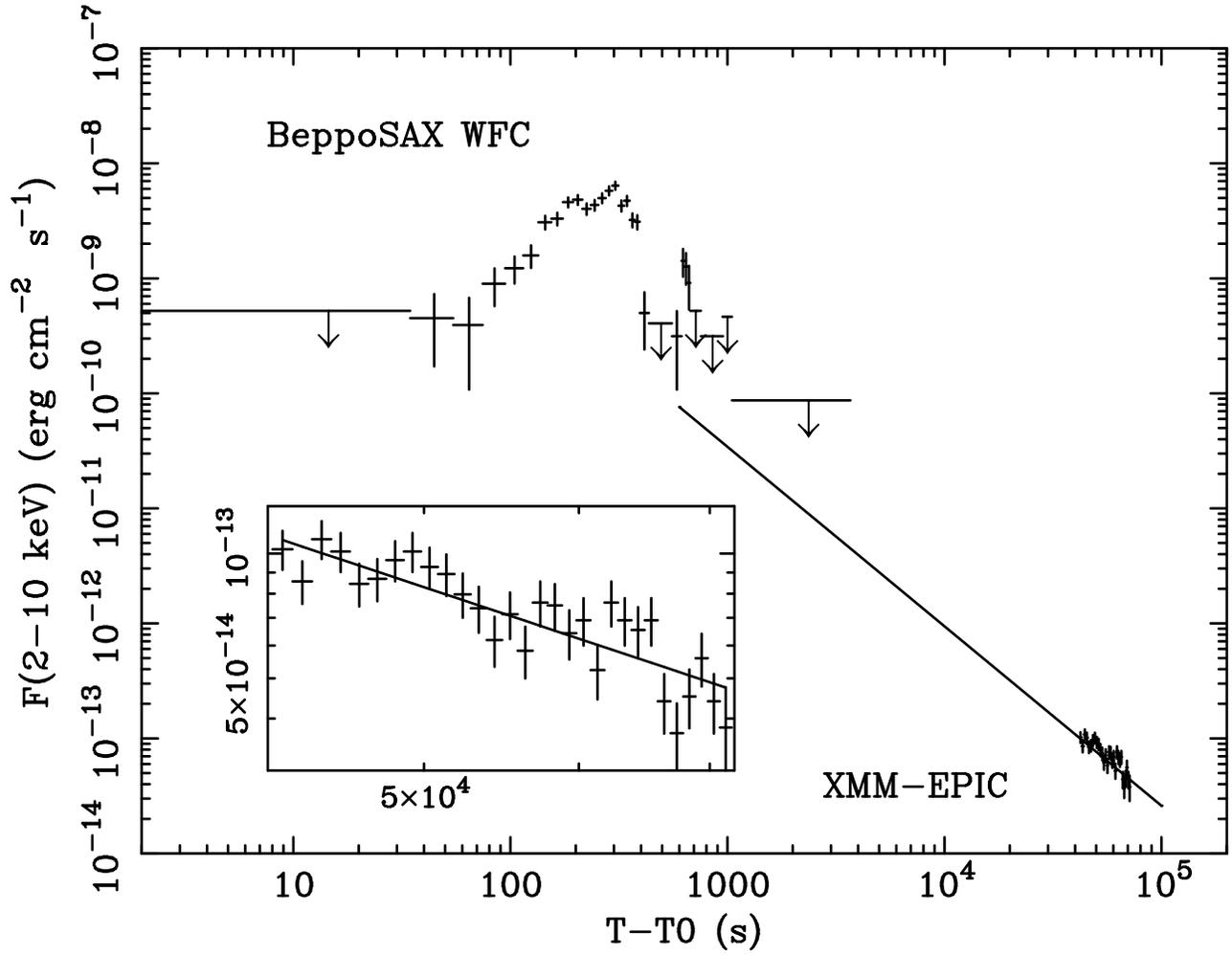}
  \caption{Light curve of the December 11, 2001 from \sax\
  WFC (from 0.1 to 5000 s), and \xmm\ pn
 (from 30.000 to 140.000 s) in the 2--10 keV range. The latter data are
 expanded in the inset. The continuous line is the best fit power law to \xmm\
 data.}
    \label{fig:011211_lc2}
\end{figure}

The prompt emission of this burst is characterized by  peculiar
features.
\begin{itemize}

\item The  event is long with a similar duration in  the gamma-ray
and X-ray ranges ($\Delta T\approx 400$s, Figure
\ref{fig:011211_lc}). On the contrary, most of the \sax\ GRBs show
a duration decreasing with energy as $\Delta T\approx E^{-0.5}$
\citep{phj+98,fac+00}. Time resolved spectral analysis shows that
the peak energy evolves from 40 keV to 12 keV
(Table~\ref{tab:gb011211}; Figure \ref{fig:011211_spe}), while a
much stronger variation of the peak energy, from $\approx$ MeV to
$\approx 10$ keV, is usually observed in long hard GRB
\citep{phj+98,fac+00}.

\item The other interesting feature is the late X-ray burst
detected in the WFC from 600 s to 700 s (Figures
\ref{fig:011211_lc} and \ref{fig:011211_lc2}). This event is
markedly distinct from the prompt phase by a gap where no emission
is detected with a 3$\sigma$ upper limit $\approx 4$ times lower
than the flux of the late burst. It contains  $\approx 4\%$ of the
fluence of the prompt event. As in the case of the November burst,
its spectrum is similar to that of the late afterglow (described
in the following section).

\end{itemize}

The possible presence of a transient absorption feature in the WFC
data is discussed in \citet{fai+04}.

\subsubsection{Follow-up observation: XMM-Newton}

The GRB was localized with the WFC at coordinates (J2000)
${\rm RA} = 11^{\rm h} 15^{\rm m} 16\fs4$, ${\rm Dec.}=-21\degr
55' 45"$ within an error box of radius $2'$. Prompt dissemination
of the coordinates \citep{g+01,g+01b} triggered follow-up
observations by several ground-based and space observatories,
including \xmm. An \xmm\ target-of-opportunity observation (TOO)
started on Dec.~12.269 UT (11.4 hours after the GRB) and lasted
for about 32 ksec. The X-ray afterglow was detected within the WFC
error box  at (J2000)
${\rm RA} = 11^{\rm h} 15^{\rm m} 17\fs9$, ${\rm Dec.}=-21\degr
56' 57\farcs5$ with an error radius of $10''$ \citep{slr+01}.

We have analyzed the EPIC (European Photon Imaging Cameras) data,
i.e. the MOS \citep{taa+01} and  pn \citep{sbd+01} CCDs , by using
different methods, to verify the consistency of the analysis
procedures. At the reduction level we used both pipeline processed
data as available from the SOC database and also independently
reprocessed raw data. Source spectra were derived by using 2
different filter options. One, that accepts only single-event
pixels, optimizes the energy resolution at expenses of source
signal. The other, that includes also double-pixel events,
increases the source signal. We found that all these different
methods lead to consistent results. In the following we present
the results derived by using the pipeline SOC-processed data with
single plus double-pixel event selection, thus optimizing the
signal-to-noise ratio. The background was estimated from
source-free regions close to the target and was fairly stable
during the whole observation.

To extract source net spectra and light curves in both pn and MOS,
we used circular regions of $30''$ radius. In the first 5 ks the
source in the pn image was located close to the edge of the gap
between CCDs.
For this period we also extracted the source counts from a circle
of $10''$, avoiding the edge of the CCD. The spectrum  was
consistent with that derived with a $30''$ radius, but  with
$\approx 40$ \% less counts.
Increasing the extraction radius from $30''$ to $40''$ did not
significantly increase the source signal, while the background
nearly doubled.  We thus adopted an extraction radius of $30''$,
that optimizes the signal-to-noise ratio. In this case the net
source count rate is $0.117\pm0.005$ cts/s and the background
contribution in the source area is 6\%.

The light curve of the X-ray afterglow is shown in Figure
\ref{fig:011211_lc2}. It follows a power law template
$F\approx{}t^{-\delta_{\rm X}}$ with $\delta_{\rm X}=1.56\pm0.25$,
consistent with the values derived by other authors from the same
data set \citep{rwo+03, jhf+03}. There is marginal evidence
($\csqn=1.25$) of excess variability in addition to the power-law
decay, with an amplitude of $\approx 10\%$ on a time scale of
hours.

Spectral analysis has been carried out by splitting the
observations in three consecutive intervals, with exposure times
of 5, 8 and 15 ks respectively. We have performed a joint fit of
p-n and MOS data. The absorption column density was kept fixed to
the Galactic one ($\NH=4.2\times10^{20}$~cm$^{-2}$), after a first
run of fit did not show any evidence for excess absorption. The
results are shown in Table~\ref{tab:gb011211}. All the spectra are
well described ($\chisqn\ltsim1$) by a power law with an average
energy index $\alpha=1.18\pm0.03$, with no evidence of variation
of the spectral index. The issue of the presence  of soft X-ray
lines has been discussed in several papers
\citep{r+02,bt03,rs03,rwo+03}. In this paper we are interested in
the spectral and temporal evolution of the continuum, so we will
not discuss this issue further.

 The optical afterglow was found 10
hours after the burst
\citep{grav01}, and its fading afterglow behaviour confirmed
\citep{bb01}; the optical flux decayed following $\approx
t^{-0.9}$ \citep{shg+01} between Dec. 12.2 and Dec. 13.3 with a
break around 2 days after the burst \citep{hbb+01}. Absorption
lines in the optical spectrum \citep{fvr+01} indicated a redshift
$z=2.14$.

\section{The origin of the X-ray precursor in the November burst}
\label{sec:precursor}

An intriguing feature that appears in the November burst is a
X-ray burst preceding the main gamma-ray pulse by $\sim30$~s. It
has roughly $2\%$ of the total fluence.  This event shows a
remarkable behaviour when compared to the proper GRB. Its spectrum
is much softer.  It is composed of a sequence of two pulses, with
the second being more intense  and softer than the first. This
trend is abruptly reversed with the onset of the gamma-ray pulse.
In a few seconds the softness ratio changes by almost two orders
of magnitude (Figure \ref{fig:011121_lc}), with the peak energy
ramping up from the keV band to above 1 MeV (Figure
\ref{fig:011121_spe}).

While this behaviour is highly suggestive of   different physical
mechanisms at work  in the precursor and in the proper burst,
 a definite conclusion  requires more observations and
theoretical efforts. In the framework of the standard scenario,
the prompt GRB phase is explained by the internal shock model
\citep{rm94,sp97}. Detailed models have been developed to describe
the temporal and spectral evolution of single and multi-pulse
bursts as produced by shocks of relativistic shells with different
Lorentz factors \citep[e.g.][]{kps97,dm98}. To our knowledge none
is  predicting a sequence made of  X-ray pulses followed by
gamma-ray pulses (see below for fireball precursors due to a
different mechanism). However, considering the  freedom of tuning
the  sequence of the ejection of the shells and their Lorentz
factor, it is not excluded that a detailed model might account for
the phenomenology observed in the November burst. Here we explore
some alternative origin for the precursor.

Thermal precursors in GRBs have been discussed in several
circumstances. Evidence  of a X-ray  precursor with a black body
spectrum  remains so far limited to a single event observed by
GINGA \citep{min+91}. In fact,  in our case, a black body model
does not provide an acceptable fit to the spectrum of the
precursor, giving $\chisq=61/27$ and $\chisq=76/27$ for the first
and second pulse respectively. This is primarily due to the
limited spectral extension of the black body, too narrow to
account for a spectrum that extends from few to hundreds of keV.
On the contrary, a power law provides a good fit (see
Table~\ref{tab:gb011121_spectra}) with a spectral index
$\alpha=1.25\pm0.1$, consistent with the spectrum observed in the
afterglow phase.

The predicted precursors can be subdivided in two classes:
fireball precursors (associated with the fireball transition from
the optical thick to the optical thin regime) and progenitor
precursors (associated with the interaction of the fireball with
the progenitor itself). Fireball precursors are predicted in
baryon loaded fireballs \citep{mr00,dm02}, magnetic outflows
\citep{lu00} and pure radiation fireballs \citep{pac86}. Fireball
precursors are expected to have a thermal spectrum and an observed
temperature ranging from several MeV for clean fireballs (either
magnetic or radiation dominated) to several tens of keV for baryon
dominated fireballs. Being $R_\gamma$ the radius at which the main
$\gamma$-ray pulse is produced, the delay between the precursor
and the main event is:
\begin{equation}
\Delta t = {{R_\gamma}\over{2c\,\Gamma_0^2}} \sim 0.2 \,
R_{\{\gamma,13\}} \, \Gamma_{\{0,2\}}^{-2} \;\; {\rm s}.
\end{equation}
where $\Gamma_0=100\ \Gamma_{\{0,2\}}$ is the Lorentz factor of
the fireball. The precursor of the November burst seems therefore
to lack all the characteristic features of a fireball precursor:
it is not thermal and it is too distant from the main event,
unless a unusually low Lorentz factor or a large radius is
inferred. The latter is the case of prompt emission produced by
external shocks \citep{rbc+02, dbc99}, but then it would be
difficult to explain the presence of the very short spikes
($\approx 1$ s, see Figure \ref{fig:011121_lc}) present in the
light curve (\cite{fmn96} but see also \cite{dm99}).

Alternative precursors associated with a shock break-out in a
massive star explosion  were originally introduced in the
framework of supernovae by \citet{kc78} and more recently
discussed for GRBs by \citet{mw99,rml02,wm03}. In this case the
precursor is due to thermal radiation coming from the shocked
stellar material that, just before the jet breaks out of the star,
is exposed on the stellar surface. Also in this case the spectrum
is expected to be of thermal nature, even though it can be
modified by interaction with the jet itself \citep{rml02}. If the
jet is optically thin at the star surface and non-thermal
particles are present, inverse Compton scattering can produce the
observed non thermal spectrum (Lazzati in prep.). Yet, the 30
seconds delay between the appearance of the precursor and the
beginning of the prompt emission in the November burst are
difficult to account for. It is possible (if not at all likely)
that the jet acceleration is not complete at the star surface, and
that the jet reaches the stellar surface with a moderate bulk
Lorentz factor and a large amount of internal energy
\citep{zwm03}. Let us call $R_\star$ the star radius and
$\Gamma_\star$ the Lorentz factor achieved by the jet at the star
surface. The delay between the precursor and the prompt emission
is then given by:
\begin{equation}
\Delta t \sim {{R_\star}\over{2c\,\Gamma_\star^2}} +
{{R_\gamma}\over{2c\,\Gamma_0^2}} \sim
3\,R_{\{\star,11\}}\,\Gamma_\star^{-2} + 0.2 \, R_{\{\gamma,13\}}
\, \Gamma_{\{0,2\}}^{-2} \;\; {\rm s}.
\end{equation}
Again it is clear that a time interval as long as 30 s calls for
non standard parameters, such as a large stellar radius, and/or a
moderate Lorentz factor.

\section{The late X-ray burst and the beginning of the afterglow}
\label{sec:rebursting}

In this section we discuss the late X-ray bursting observed in the
December and November bursts at $t=600$ s and $t=240$ s
respectively. Correcting for the redshift, they correspond to a
similar $t\approx 185$~s.

The similarity between the spectrum of these events  with that of
the afterglow observed during the follow-up observations with
\xmm\ and \sax\ is particularly compelling. Furthermore in the
November burst  the spectrum of the late X-ray burst was markedly
softer than that of the preceding emission. This behaviour is
strikingly similar to that observed in other bursts, and the only
difference is the onset of the X-ray bursting \citep{fac+00}. For
example in GRB970228 soft  bursting activity started 35 sec after
the main hard pulse. The spectrum was similar to that observed in
the late afterglow observations \citep{fcp+98}. This has been
associated with the transition from the prompt hard phase to  the
afterglow \citep[][hereafter SP99]{fac+00,sp99a}. It is therefore
tantalizing to identify the late X-ray bursting as the onset of
the external shock. Such an interpretation and the consequences it
bears are not straightforward.

The onset of external shocks depends on the dynamical conditions
of the fireball and, in particular, two regimes can be identified
depending on the ``thickness'' of the fireball (SP99). In the
regime of thin shell, the reverse shock crosses the shell before
the onset of the self-similar solution (i.e. when an ISM mass
$m=M_0/\Gamma_0$ is collected, $M_0$ and $\Gamma_0$ being the rest
mass and asymptotic Lorentz factor of the fireball respectively).
As a consequence, the onset of the afterglow coincides with the
deceleration time:
\begin{equation}
t_{\rm{aft}}=\left\{
\begin{array}{cc}
\left({{3\,E}\over{32\pi\,n\,m_p\,c^5\,\Gamma_0^8}}\right)^{1/3}
& {\rm ISM} \\
\frac{E}{4\pi\times10^{12}\,A_\star\,c^3\,\Gamma_0^4} & {\rm Wind}
\end{array} \right.
\end{equation}
where $E$ is the kinetic energy of the fireball, $n$ the density
of the ISM, $A_\star=\frac{\dot M_w/10^{-5} M_{\odot} {\rm
yr^{-1}}}{v_w/10^3 {\rm km\ s^{-1}}}$,  $M_w$ is the mass-loss
rate and $v_w$ the wind velocity.

In this case the peak of the afterglow is well separated from the
prompt phase (SP99), as long as the prompt phase is identified
with internal shocks. Moreover, the evolution of the afterglow
after the peak is well described by a power-law decay, if the time
is measured starting from the explosion time, very well
approximated by the time at which the first prompt phase photons
are collected. In the case of a thick shell, the reverse shock has
not crossed the shell when the critical mass $m=M_0/\Gamma_0$ has
been collected, and therefore the external shock keeps being
energized for a longer time. The peak of the afterglow emission
therefore coincides  with the shell crossing time of the reverse
shock:
\begin{equation}
t_\Delta=\Delta/c \approx T
\end{equation}
where $T$ is the duration of prompt phase and $\Delta$ the
thickness of the shell. It is immediately obtained that the early
afterglow emission is mixed with the late GRB one. Also, the
afterglow decay will be well described with a single power-law
only if the time is measured starting from the time at which the
inner engine turns off, roughly coincident with the GRB duration.
Mathematically, a shell is defined as thick if (SP99):
\begin{eqnarray}
T_2 \left({{n\,}\over{E_{52}}}\right)^{1/3}\Gamma_{\{0,2\}}^{8/3}
&>& 2.9 \qquad {\rm ISM} \\ \nonumber
\frac{T_2\,A_\star\,\Gamma_{\{0,2\}}^4}{E_{52}} &>& 0.006
\qquad {\rm Wind}
\label{eq:thin}
\end{eqnarray}
 and thin otherwise.

\begin{figure}
\centering
\includegraphics[width=0.5\textwidth,origin=c,angle=270]{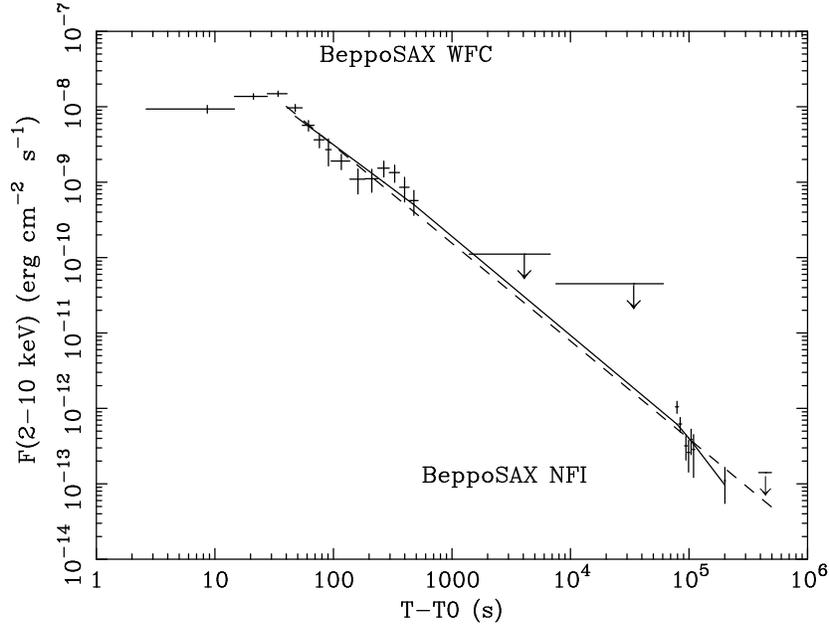}
\includegraphics[width=0.5\textwidth,origin=c,angle=270]{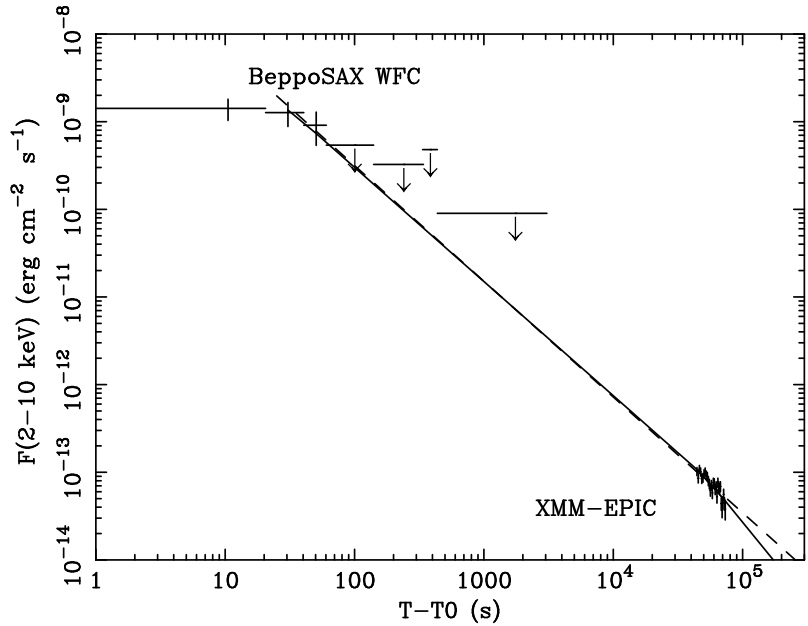}
\caption{Light curves of the November 21, 2001 burst (upper
figure) and the December 11, 2001 (lower figure) in a log-log
scale derived by setting $t_0$ at the onset of  the late X-ray
burst ($t_0=240$~s for the November burst and $t_0=614$~s for the
December burst). The dashed line is the best fit power law decay
model, while the continuous line is the best fit jet model (see
text).} \label{fig:aft_reb}
\end{figure}

In our cases, for $n=1$ and $\Gamma_{\{0,2\}}=1$, the thin shell
condition is satisfied for both bursts, while for $A_\star=0.003$
(see below) and $\Gamma_0=100$ the thin shell is also satisfied
for the wind best fit case for the November burst. However the
strong dependence on $\Gamma$ of Eq.~\ref{eq:thin} allows for
solutions in which a thick shell approximation is valid. We now
present some evidence supporting this case. The decay part of the
late X-ray burst cannot be connected to the 1-day afterglow
emission with a single power-law  $(t-t_0)^{-\delta_{\rm X}}$ when
$t_0$ corresponds to the GRB onset (e.g.
Figure~\ref{fig:011211_lc2}). On the contrary, this is the case
(Figure \ref{fig:aft_reb}) when $t_0$ is set equal to the onset of
the late X-ray burst.
The best fit  slopes are $\delta_{\rm X}=1.33\pm0.07$
($\chisq/dof= 39/30$) and $\delta_{\rm
X}=1.29\pm0.04$($\chisq/dof= 31/16$) for the December and November
bursts respectively . The errors of the decay slopes include
systematic errors derived by changing $t_0$ of $\pm10$~s.

This observation seems therefore to favor a situation in which the
shell is thick (and therefore either a dense medium or a
relatively large Lorentz factor $\Gamma>100$). The fact that the
afterglow peak is well separated from the $\gamma$-ray phase would
imply a long energy release by the inner engine, lasting longer
than the observed $\gamma$-ray phase. This can be obtained if the
efficiency of conversion of the kinetic energy into $\gamma$-rays
decreases with time (e.g. due to a smaller dispersion of the
Lorentz factor).

Alternative explanations would require  an inhomogeneous fireball
or external medium. In this case the late X-ray burst would not be
identified with the afterglow onset, but with an emission bump
overlaid on the regular afterglow decay.   In the X-ray regime,
inhomogeneities in the external medium can hardly affect the light
curve, since the relevant electrons are in the fast cooling regime
\citep{lrc+02}. Inhomogeneities in the fireball itself and/or a
re-energization of the shock by a delayed shell are both possible
explanations. The latter hypothesis was proposed by \citet{pmr98}
to explain the later and longer re-bursting event observed in
GRB970508 \citep{paa+98}.   A major re-energization or hot spot is
however necessary in order to obtain such a prominent bump in the
light curve.

\section{A possible break in the X-ray light curves} \label{sec:break}


 According to
\citet{gks+03}, the optical light curve of the November burst
shows a break at $1.2\pm1.0$ days, i.e. slightly after the
beginning of the BeppoSAX NFI observation. The post-break optical
slope $\delta_{\rm O2}=2.44\pm0.38$ is consistent with the steep
slope derived by fitting the data of the BeppoSAX NFI
alone\footnote{This is not sensitive to variation of $t_0$ of
$\approx 100$~s} ($\delta_{\rm X}=3.8\pm1.9$). In the December
burst \citet{jhf+03} found a break in the optical curve   at
$t=1.56\pm0.02$ days with a post-break slope of $\delta_{\rm
O2}=2.11\pm0.07$.

We have therefore fitted again the early and late afterglow X-ray
light curves employing the empirical formula \citep{bhr+99}
\begin{equation}
F_{\rm X}(t)= 2^{1/n}F_{\rm X}(t_b)
\left[\left(\frac{t}{t_{b}}\right)^{n\delta_{\rm X1}}+
\left(\frac{t}{t_{b}}\right)^{n\delta_{\rm X2}}\right]^{-1/n},
\label{eq:break}
\end{equation}
where the parameter $n$ describes the sharpness of the transition.
To limit the number of free parameters, we have analyzed the cases
of $n=10$ (sharp transition) and $n=1$ (shallow transition). In
addition we have fixed the post break transition $\delta_{\rm X2}$
to the values derived in the optical band. The results are shown
in Figure \ref{fig:aft_reb}.

For the November burst we derive  $\delta_{\rm X1}=(1.22\pm0.09)$
and $\delta_{\rm X1}=(1.15\pm0.09)$ for a sharp and shallow
transition respectively. The corresponding  break times are
$t_b=(0.7\pm0.3)$day and $t_b=(0.7^{+1.0}_{-0.4})$ day, consistent
with  optical results.  This model gives $\chisq/dof=23.5/15
(25.7/15)$ with a marginal improvement ($97\%$) with respect to
the simple power law.

For the December burst we have limited the analysis to the case of
a shallow transition, because the optical break takes place after
the end of the \xmm\ observation.  We derive
$t_b=(0.8^{+0.5}_{-0.3})$ days, consistent with the value derived
from optical data, and $\delta_{\rm X1}=1.22\pm0.06$.  This model
gives $\chisq/dof=34.9/29$, with a very marginal ($93\%$)
improvement with respect to the simple power law.

We conclude that, although the X-ray light curves are consistent
with the presence of a break as found in the optical band,  the
statistical evidence  is not compelling.

\section{The fireball evolution: constant density and wind
environments} \label{sec:011121_evolution}

By using the parameters of the X-ray spectral and temporal
evolution  and taking into account the optical behavior, we now
show that the afterglow emission in the December burst is
consistent with an expansion in a constant density medium. On the
contrary a wind is required for the November burst. This analysis
is carried out for $t\lesssim t_b$, when the expansion is
described by a spherical fireball.

We recall that, according to the fireball model, the temporal and
spectral slopes are linked through relationships that depend upon
the kind of the expansion (spherical or jet), and the density of
the medium (uniform or wind) \citep[e.g.][]{sph99,cl00}. We
computed the so-called closure relationships in all the relevant
cases considering at first only the X-ray spectral and temporal
slopes.  In both bursts the only solution consistent with the data
is the case of spherical expansion into either ISM or wind, for a
cooling frequency below the X-ray range. For the December burst
the  corresponding closure relation is $C=\delta_{\rm
X}-3/2\alpha_{\rm X}+1/2=0.08\pm0.09$, while for the November
burst it is $C=0.06\pm0.2$ or $C=-0.38\pm0.6$, adopting the
spectra index of the late X-ray burst or of the late afterglow
respectively.


The degeneracy of the solution   wind vs ISM can be resolved by
comparing the temporal decay slopes above and below the cooling
frequency, i.e. in X-rays and optical. In a wind the temporal
evolution in X-rays should be shallower than in the optical
($D=\delta_{\rm X}-\delta_{\rm O}=-0.25$), while the reverse holds
for an expansion in an ISM ($D=+0.25$).

\begin{figure}
\centering
\includegraphics[width=\textwidth,origin=c,angle=0]{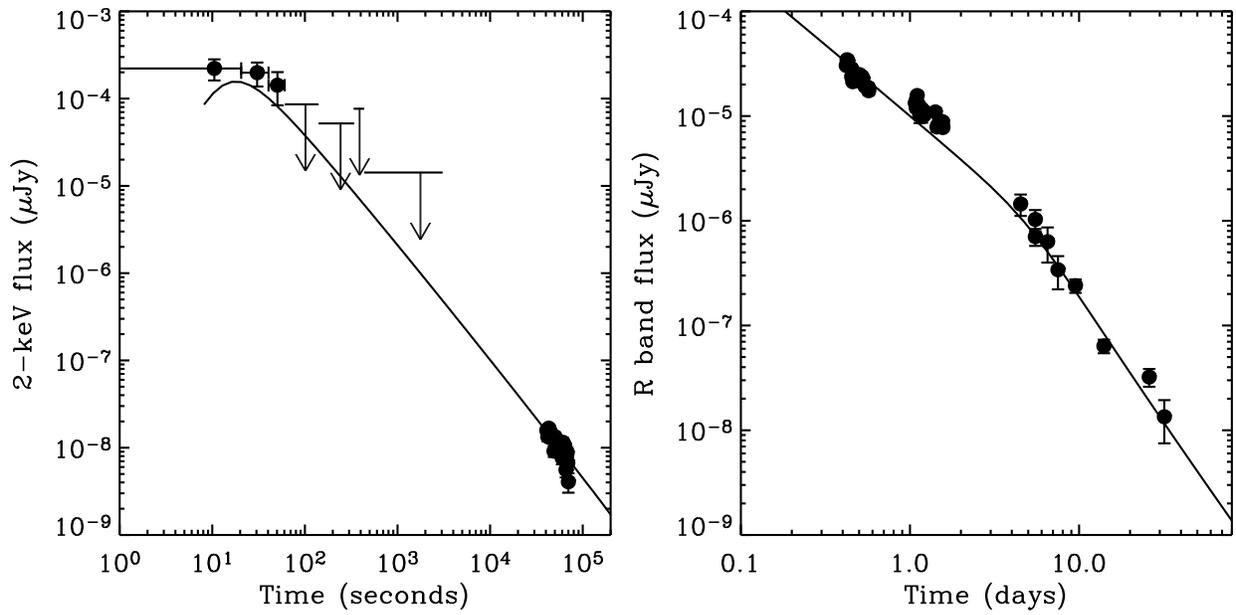}

\caption{X-ray (2 keV, left panel) and optical (R band, right
panel) light curves of the afterglow of the December 11, 2001
burst. The solid line is an example model for an external shock
emission in a uniform medium (see text).
}

 \label{fig:fit011211}
\end{figure}

In the December burst $\delta_{\rm O}=0.95\pm0.02$ \citep{jhf+03}
vs $\delta_{\rm X}=1.3\pm0.07$, that gives $D=0.4\pm0.07$,
consistent only with the ISM case. The value of the electron index
distribution $p\approx2.4$ derived from the temporal slopes
implies optical and X-ray spectral slopes of 0.7 and 1.2
respectively, in good agreement with the observed values
$\alpha_{\rm O}=0.66\pm0.13$ \citep{gsw+03} and $\alpha_{\rm
X}=1.18\pm0.03$.
Guided by these results, we have applied a detailed modelling of
the afterglow using the prescription given e.g. in \citet{pk00}.
In Figure \ref{fig:fit011211} we present a  model with $p=2.45$,
total energy $E_{53}=0.9$, $\epsilon_e=0.0025$ and
$\epsilon_b=0.01$ (the fraction of energy in relativistic
electrons and magnetic field respectively), density $n=3$
cm$^{-3}$ and jet opening angle $\theta_j=8^{\circ}$.

For the November burst only the wind case is compatible with the
data, because $\delta_{\rm O}=1.66\pm0.06$ \citep{pbr+02} vs
$\delta_{\rm X}=1.29\pm0.04$, yielding $D=-0.37\pm0.07$. This
gives $p\approx2.5$, with predicted spectral slopes in the optical
(0.75) and X-rays (1.25) consistent with the observed values
$\alpha_{\rm O}=0.76\pm0.16$ \citep{pbr+02} and $\alpha_{\rm
X}=1.15\pm0.15$ (Table~\ref{tab:gb011121_spectra}). In the next
section we present a detailed afterglow modelling.

\section{The late radio to X-ray afterglow data of the November burst: a
wind termination shock?} \label{sec:011121_aft}

Following the result presented in the previous section, we have
carried out a detailed modelling of  the broad band afterglow data
of the November burst.

\begin{figure}
\centering
\includegraphics[width=\textwidth,origin=c,angle=0]{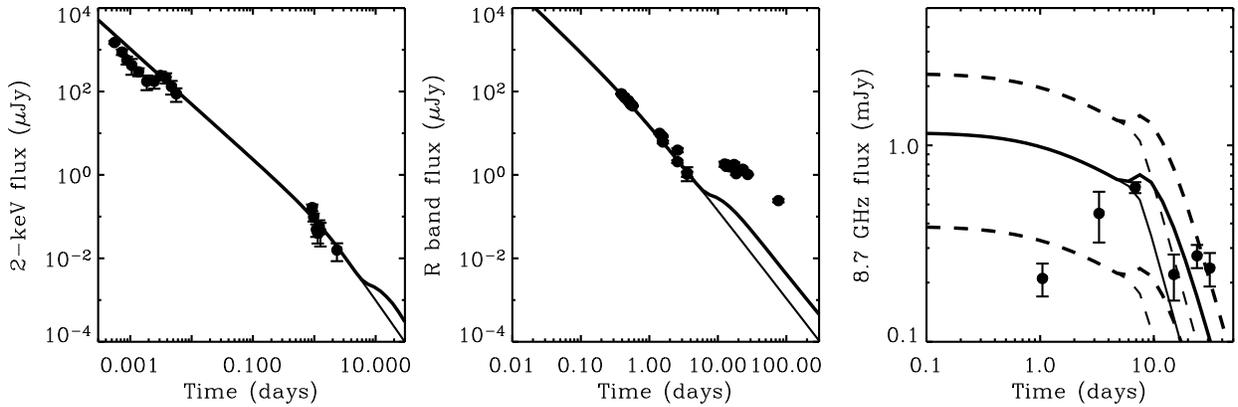}

\caption{X-ray (2 keV, left panel), optical (R Band; central
panel) and Radio (8.7 GHz; right panel) light curves of the
afterglow of the November 21, 2001 burst. Note that the optical
light curve after 10 days is likely dominated by a SN contribution
\citep{gsw+03, gks+03}. The thin solid line is the model
reproducing the optical, X-ray and early radio data in a wind
environment. Such a model cannot reproduce the late time radio
data, that can be accounted for (thick solid line) only if a
different scaling for the density with radius is assumed at
$r>3$~pc. Thin and thick dashed line give the corresponding range
of uncertainty due to ISS.} \label{fig:fit}
\end{figure}

We have first fitted the data at 1 day, i.e. when the expansion is
described by a spherical fireball. Data points are the X-ray,
optical fluxes
 and the
radio flux at 8.7 GHz, including the large uncertainty due to
interstellar scintillation (ISS) \citep{pbr+02}. The total energy
has been fixed to the observed value $E_{53}=0.28$ and $p=2.5$, as
derived in the previous section.
  With these
constraints we find $A_*\sim0.003$, $\epsilon_e\sim0.01$ and
$\epsilon_B\sim0.5$. This solution is shown in the central and
leftmost panel of Figure~\ref{fig:fit}. Interestingly, if we relax
the constraint on $E_{53}$, allowing for a small efficiency of the
internal shock phase, no acceptable solution can be found,
pointing to a large efficiency for the $\gamma$-ray production.


However, if we blindly extrapolate the above mentioned model to
the radio band assuming a jet evolution after 1.3 days, we cannot
account for the late time radio measurements \citep{pbr+02}, which
are far brighter than the prediction, even considering the ISS.
This is due to the steep decrease of the model flux at around 10
days, when the injection frequency enters the radio band. This
behaviour cannot be changed by any other choice of model
parameters without violating the constraints on  the injection and
cooling frequencies imposed by optical and X-ray data. They
require the injection frequency to be below the optical band and
the cooling frequency to be in between the optical and X-ray
bands.
The extrapolated
model is shown in the rightmost panel of Figure~\ref{fig:fit} with
a thin solid line; thin dashed lines account for possible flux
variations induced by ISS \citep{w98}.

It should be kept in mind, however, that the $r^{-2}$ scaling for
the density cannot hold out to very large radii, since the stellar
wind does not expand in vacuum but rather in a dense molecular
cloud. The wind interaction with the molecular cloud gives rise to
a complex discontinuity, and shock structure recently discussed by
\citet[]{clf04}[ see their Figure~1]. Given the paucity of data,
we account for this by assuming a uniform density for the ISM
after a given radius $r_w$. We find that the radio data at
$t>20$~days can be reproduced if the density becomes uniform at a
radius $r_w=3$~pc with a low density value $n=10^{-5}$~cm$^{-3}$.
The model is shown in all the panels of Figure~\ref{fig:fit} with
a thick solid line. In the radio band panel, the ISS fluctuation
range is shown with thick dashed lines.  Since the radio
flattening occurs after the jet break, the exact value of $r_w$
depends on the details of the jet sideways expansion. The quoted
number holds for a non sideways expanding jet, while smaller $r_w$
are relevant in an expanding case.

Following \citet{clf04} we find that the wind structure is
consistent with what expected from a fast light Wolf-Rayet wind
which expands in -- and is confined by -- a high pressure
environment. In this case a large constant density region is
produced in the shocked wind.  This and the low wind density
inferred from our afterglow modelling, give a wind with
$\dot{M}\sim10^{-7}$~$M_\odot$~y$^{-1}$ and
$v_w\sim3\times10^8$~cm~s$^{-1}$.

\section{The origin of the X-ray absorber in the November burst}
\label{sec:absorber}
 A significant absorption with a column
density $\NH=(7\pm2)\times 10^{22}$ cm$^{-2}$ is detected during
the prompt emission until $t=25$ s (Table~\ref{tab:fit011121}).
After this time, the column density   decreases to a value
consistent with zero, although some of the upper limits are still
marginally consistent with the value observed in the early phase.
An overall decrease of the column density  is expected if the
absorbing gas is contained in a region of a few pc around the
burst, due to ionization by hard photons \citep{pl98,lp02}.  We
have at first explored the possibility that this absorption is
produced by the wind. A simple general formula is derived in this
case:
\begin{equation}
N_{\rm H,22}=3\frac{A_*}{r_{13}}, \label{eq:nh}
\end{equation}
where $\NH=10^{22}N_{\rm H,22}$ cm$^{-2}$ is the column density of
the wind from a radius $r=10^{13}r_{13}$ cm to infinity. For the
typical radius at which the prompt emission is produced in the
internal shock scenario, $r_{13}=1$ and $A_*=0.003$  eq.
\ref{eq:nh} gives $N_{\rm H,22}= 10^{-2}$. The column density of
the wind is thus much lower than observed.
 We conclude that the
absorbing medium is external to the fireball region, and it could
be associated  with a star-forming region embedding the GRB.

\section{Conclusions}
\label{sec:conclusions}

 In this paper we have presented the spectral
and temporal evolution of the prompt and afterglow emission of two
bursts, the events of November 21, 2001 (GRB011121) and of
December 11, 2001 (GRB011211). The results have relevant
implications on the environment and the progenitor.

Both events show a late X-ray burst taking place hundreds of
seconds after the prompt emission. This phenomenon might also be
present in other bursts, particularly in some of the long duration
GRB identified in an off-line analysis of BeppoSAX WFC data
\citep{ihk+04}. For the bursts presented in this paper we find
that the spectrum of the late X-ray burst is substantially softer
than the prompt emission and remarkably similar to the power law
observed in the afterglow at later times. This behaviour has been
observed in several other bursts \citep{fac+00}, but on shorter
time scales,
 and it is attributed to the transition from the prompt emission
to the early afterglow in the framework of the internal-external
shock scenario (SP99). These two phases are typically separated by
gaps with little or undetected emission, but there are cases where
the two overlap \citep{pfg+02,sdp+04}. The tail of the early
afterglow emission usually lies on the backward extrapolation of
the power law decay observed a day after the burst. In the
November and December burst we find that the tail of the late
X-ray burst
 and the  afterglow at 1 day are connected with a single power law, but only if
 $t_0$ is  coincident with the onset of the late burst.
 This is what is expected in the case of a thick fireball,
 where the afterglow decay is
described by a single power law only if the time is measured
starting from the time at which the inner engine turns off (SP99).
  We are
thus led to the conclusion that the late X-ray burst represents
the beginning of the afterglow.
 The fact that the afterglow peak is
delayed from the $\gamma$--ray phase would imply a long energy
release from the inner engine, lasting longer than the observed
prompt phase. This can be obtained if the efficiency of conversion
into $\gamma$--rays decreases with time (e.g. due to a smaller
dispersion of the Lorentz factor in the internal shock scenario).

An intriguing feature observed in the November burst is a  X-ray
burst
 preceding by $\approx 30$ s  the hard pulse of the
$\gamma$-ray burst. Similar events have been observed in few other
bursts \citep{ihp+99,faa+00}. It contains roughly 2\% of the
fluence of the main event. The spectrum is not consistent with a
black body and is well fitted with a power law with energy index
of 1.2 extending from 2 to 700 keV. The origin of this feature is
puzzling.  Precursors associated with shock break-out at the
surface of the star have been discussed in the framework of the
association of GRBs with massive star explosions. These events are
expected to be of thermal nature, although a modification to a
non-thermal spectrum can be obtained through the interaction with
a non-thermal component in the jet at the stellar surface. In this
framework, the delay between the precursor and the prompt emission
would require that the jet acceleration is not complete at the
star surface, and that the jet reaches the stellar surface with a
relatively small bulk Lorentz factor.

The spectral and temporal behaviour of the X-ray afterglow and the
time decay of the optical afterglow indicate that the fireball
expands in a constant density environment in the December burst.
From broad band afterglow modelling we derive  $n=3$ cm$^{-3}$,
fireball total energy $E_{53}=0.9$, $\epsilon_e=0.0025$ and
$\epsilon_b=0.01$ and jet opening angle $\theta_j=8^{\circ}$. On
the contrary in the November burst a fireball expansion in a wind
 is clearly revealed by a X-ray decay (temporal slope
$\delta_{\rm X}=1.29\pm0.04$) slower than the optical
($\delta_{\rm O}=1.66\pm0.06$). Broad-band modelling of radio to
X-ray data at 1 day requires a wind with a rather low density $A_*
\approx 0.003$, a total isotropic energy similar to that observed
in gamma-rays (indicating a high efficiency for $\gamma$-ray
production), $\epsilon_e\approx 0.01$, $\epsilon_B\approx 0.5$.
The X-ray data are consistent with (but do not require) a break at
1.3 days, as suggested by optical data \citep{gks+03}. However,
the late-time radio data \citep{pbr+02} fall above the
extrapolation of a collimated fireball in a wind. We find that
this discrepancy can be solved if the density becomes uniform with
$n=10^{-5}$ cm$^{-3}$ at a radius $r_w=3$ pc. This wind structure
is consistent with that expected from a Wolf-Rayet wind which
expands in -- and is confined by -- a high pressure environment
\citep{clf04} as that expected in a star-forming region. In this
case a large constant density region is produced in the shocked
wind. For the November burst we derive a wind with
$\dot{M}\sim10^{-7}$~$M_\odot$~yr$^{-1}$ and
$v_w\sim3\times10^8$~cm~s$^{-1}$.

Finally, a significant absorption with a column density
$\NH=(7\pm2)\times 10^{22}$ cm$^{-2}$ is detected during the
prompt emission of the November burst until $t=25$ s. After this
time, the decreases to a value consistent with zero, although some
of the upper limits are still marginally consistent with the value
observed in the early phase. A  decrease of the column density on
a time scale similar to that observed is expected if the absorbing
gas is contained in a region of a few pc around the burst, due to
ionization by hard photons \citep{pl98,lp02}. Since the column
density in the wind is much lower than observed,
 we conclude that the
absorbing gas   could be associated  with the medium of the
star-forming region embedding the GRB, that is also likely
responsible for the termination shock in the wind structure
discussed above.

\acknowledgments  This paper is based on observations made with
\sax\ , a program of the Italian space agency (ASI) with
participation of the Dutch space agency (NIVR), and with
XMM-Newton, an ESA science mission with instruments and
contributions directly founded by ESA member states and USA
(NASA).  LP, EC, MF acknowledge support by EU through the EU FPC5
RTN ``Gamma-Ray Bursts: an enigma and a tool"

\normalsize
%


\clearpage
\clearpage

\begin{deluxetable}{lcc}
\tablecolumns{3} \small \tablewidth{0pt} \tablecaption{ Peak flux
$F$ and fluence S of the prompt emission of the November 21, 2001
burst (011121) and of the December 11, 2001 burst (011211).}
\tablehead{\colhead{ } & \colhead{011121} & \colhead{011211} }
\startdata
    $F_{40-700}$ (erg cm$^{-2}$ sec $^{-1}$)    &  $7.3\times10^{-6}$    &  $5\times 10^{-8}$        \\
    $F_{2-26}$   \hspace{1.5cm} ``      &  $6\times10^{-7}$    &  $1.4\times 10^{-8}$  \\
    $F_{2-10}$   \hspace{1.5cm} ``      &  $3\times10^{-7}$    &  $0.7\times 10^{-8}$  \\
    S$_{40-700}$   (erg cm$^{-2}$)        &  $1\times10^{-4}$     &  $5.1\times 10^{-6}      $  \\
    S$_{2-26}$     \hspace{0.8cm}  ``        &  $1.4\times10^{-5}$     &  $2.2\times 10^{-6}$  \\
    S$_{2-10}$     \hspace{0.8cm} ``      &  $0.7\times10^{-5}$    &  $1.1\times 10^{-6}$  \\
    $\frac{S_{2-26}}{S_{40-700}}$& 0.14 &0.5 \\
    $\frac{F_{2-26}}{F_{40-700}}$& 0.08 &0.3 \\
\enddata
\tablecomments{Subscripts indicate the energy range, expressed in
keV. Peak fluxes are calculated with a bin size of 8 seconds.}
\label{tab:prompt}
\end{deluxetable}


\begin{deluxetable}{ccccccccc}
\tabcolsep0.02in \footnotesize
\tablewidth{0pt}
 \tablecaption{X-ray spectral fits to \sax/WFC+GRBM and MECS+LECS of the November 21, 2001 burst.
\label{tab:fit011121}}
 \tablehead { \colhead{slice} & \colhead {T$_1$;T$_2$} &
\colhead {Instrument} & \colhead {$F_{\rm 2-10\; keV}$}  &
\colhead{$F_{\rm 40-700\; keV}$} & \colhead{$\NH$} &  \colhead {$\alpha$}
&
\colhead{$\chisq/\nu$} \\
 \colhead{} &  \colhead{s}& \colhead{} &  \colhead{\fu} & \colhead{\fu}&
 \colhead{$10^{22}$~cm$^{-2}$} & \colhead{}
& \colhead {}}
\startdata
Precursor 1 &$-28;-16$ & WFC+GRBM& $1.3\times 10^{-8}$ & $2.8\times 10^{-8}$  &$3.0_{-3.0}^{+3.5}$ & $1.0\pm0.11$ &  25.5/26  \\
Precursor 2 &$-16;-6$ & WFC+GRBM& $4.1\times 10^{-8}$ & $3.5\times 10^{-8}$  &$4.0\pm3.0$ & $1.25\pm0.10$ &  18.9/26  \\
 Rise & $-6;9$ & "& $8.2\times 10^{-8}$  & $1.8\times 10^{-6}$ &$4.5\pm2.5$ & $0.37\pm0.03$ &  18.5/27  \\
$\gamma$-peak &9;13 & "  & $1.2\times 10^{-7}$  & $6.6\times 10^{-6}$ &$3.7_{-3.7}^{+5.3}$ & $0.12\pm0.05$ &  31.1/27  \\
intermediate & 13;18  & "  & $1.6\times 10^{-7}$  & $4.6\times 10^{-6}$ &$9.5\pm3.5$ & $0.33\pm0.04$ &  19.0/27  \\
X-ray peak  &18;24  & "  & $2.2\times 10^{-7}$  & $2.2\times 10^{-6}$ &$6.8\pm2.2$ & $0.58\pm0.04$ &  21.6/27  \\
quick tail& 24;31 & "  & $1.2\times 10^{-7}$  & $1.0\times 10^{-6}$ &$<1.0$ & $0.58\pm0.03$ &  40.6/27  \\
slow tail&31;45  & "  & $4.0\times 10^{-8}$  & $4.8\times 10^{-7}$ &$<1.0$ & $0.47\pm0.04$ &  33.5/26  \\
slower tail&45;150  & "  & $7.9\times 10^{-9}$  & $8.0\times 10^{-8}$ &$5\pm4$ & $0.57\pm0.04$ &  20.1/26  \\
intermediate&150;239  & "  & $2\times 10^{-9}$  & $7\times 10^{-9}$ &$0$ fix & $0.8\pm0.15$ &  36.7/27  \\
rebursting &239;308  & "  & $8.9\times 10^{-9}$  & $9.3\times 10^{-9}$ &$1_{-1}^{+3}$ & $1.15\pm0.15$ &  32.0/26  \\
slow tail &308;716  & WFC  & $1.2\times 10^{-9}$  & - &$5_{-5}^{+32}$ & $1.3_{-0.6}^{+1.3}$ &  22.8/25  \\
afterglow &(76;120) $10^4$  & LECS+MECS  & $4\times 10^{-13}$  & - & $<10$ & $1.6\pm0.7$ &  11/18  \\
\enddata
\tablecomments{$\alpha$ is the energy spectral index. $\NH$ in the
rest frame of the burst. Errors are at $90\%$ confidence level for
one interesting parameter. Fluxes are derived from the best fit
model.} \label{tab:gb011121_spectra}
\end{deluxetable}
%
%
%
%
%
%
%

\begin{deluxetable}{ccccccccc}
\tabcolsep0.03in
\tablewidth{0pt}
 \tablecaption{X-ray spectral fits to \sax/WFC+GRBM and \xmm-EPIC
 data of the December 11, 2001 burst.} \tablehead { \colhead {T$_1$; T$_2$} &
\colhead {Instrument} & \colhead {$F_{\rm 2-10\; keV}$}  &
\colhead{$\NH$} &  \colhead {$\alpha$} & \colhead{$\beta$} &
\colhead{$E_0$} &
\colhead{$\chisq/\nu$} \\
  \colhead{s}& \colhead{} &  \colhead{(\fu)}& \colhead{($10^{22}$~cm$^{-2}$)} & \colhead{}
& \colhead{} & \colhead{(keV)}
& \colhead {}} \startdata
0;300 & WFC+GRBM$^{1,2}$& $2.4\times 10^{-9}$  & - &$0.2\pm0.2$ &
$1.3_{-0.2}^{+2.0}$ &
$50_{-20}^{+60}$ &  47.9/57  \\
300;400  & WFC+GRBM$^{1,2}$ & $3.5\times 10^{-9}$   & - & " & " & $15_{-10}^{+35}$ & "\\
575;675 & WFC+GRBM$^{1}$ & $9.5\times 10^{-10}$ & - & 0.26 fix  &
1.3 fix & $10_{-6}^{+25}$  &  26.0/28\\
0;300 & WFC+GRBM$^{}$& $2.4\times 10^{-9}$  & $40\pm20$ & $0.79\pm0.05$ & - & - &  25.8/27  \\
300;400  & WFC+GRBM$^{}$ & $3.5\times 10^{-9}$   & $60\pm30$ & $1.18\pm0.12$ & - & - & 23.5/27\\
575;675 & WFC+GRBM$^{}$ & $7\times 10^{-10}$ &  $40_{-40}^{+80}$ & 1 fix  & - & - &  28.6/28\\
(41;46) $10^4$ &  EPIC pn+MOS$^{3}$ & $1.1\times 10^{-13}$   & - & $1.09\pm0.07$ & -& - &59.9/62\\
(47;55) $10^4$ & EPIC pn+MOS$^{3}$& $8.6\times 10^{-14}$  & - & $1.20\pm0.05$ &- &- &  67.7/72\\
(55;71) $10^4$ & EPIC pn+MOS$^{3}$& $4.2\times 10^{-14}$   & - & $1.17\pm0.04$  &- &-  & 88.2/86\\
\enddata
\tablecomments{$\alpha$ and $\beta$ are energy spectral indexes.
$\NH$ in the rest frame of the burst. Errors are at $90\%$
confidence level for one interesting parameter. $^{1}$Band model
with absorption column density fixed to the Galactic value.
$^{2}$Joint fit to the (0--300)~s and (300--400)~s data sets with
$\alpha$ and $\beta$ linked. $^{3}$Power-law model with absorption
column density fixed to the Galactic value.} \label{tab:gb011211}
\end{deluxetable}

\clearpage

\end{document}